\documentstyle[epsf]{mn}

\newcommand{\etal}{et~al.}

\newcommand{\ionhy}{H{\sc ii}}
\newcommand{\molhy}{$\mbox{H}_{2}$}

\newcommand{\water}{$\mbox{H}_{2}\mbox{O}$}
\newcommand{\methanol}{$\mbox{CH}_{3}\mbox{OH}$}

\newcommand{\ammonia}{$\mbox{NH}_{3}$}
\newcommand{\transa}{$5_{1}\rightarrow6_{0}\mbox{~A}^{+}$}

\newcommand{\recomb}{H91$\alpha$}
\newcommand{\kms}{$\mbox{km~s}^{-1}$}

\newcommand{\micron}{\mbox{$\mu$m}}
\newcommand{\mjb}{mJy beam$^{-1}$}

\def\epsout #1 {
  \centering
     \leavevmode\epsffile{./#1}
}

\title[Continuum emission associated with 6.7-GHz methanol masers]
      {Continuum emission associated with 6.7-GHz methanol masers}
\author[S.P. Ellingsen, R.P. Norris \& P.M. McCulloch]
       {S.P. Ellingsen$^{1}$, 
	R.P. Norris$^{2}$,
	and P.M. McCulloch$^{1}$ \\
	$^{1}$Physics Department, University of Tasmania, GPO Box 252C, 
	      Hobart, TAS 7001.\\
	$^{2}$Australia Telescope National Facility, CSIRO, PO Box 76,
	      Epping, NSW 2121.}
\date{Received dd Month Year; in original form dd Month Year}
\pagerange{\pageref{firstpage}--\pageref{lastpage}}
\pubyear{1995}

\begin{document}

\label{firstpage}

\maketitle

\begin{abstract}

We have used the Australia Telescope Compact Array (ATCA) to search
for continuum emission toward three strong 6.7-GHz methanol maser
sources.  For two of the sources, G339.88-1.26 and NGC~6334F
(G351.42+0.64), we detect continuum emission closely associated with
the methanol masers.  A further three clusters of masers showed no
radio continuum emission above our sensitivity limit of 1-5~mJy.  We
find the position of the 6.7-GHz methanol masers in G339.88-1.26 to be
consistent with the hypothesis that the masers lie in the
circumstellar disc surrounding a massive star.  We also argue that one
of the clusters of methanol masers in NGC~6334F provides indirect
observational support for the circumstellar disc hypothesis.

\end{abstract}


\section{Introduction}

Maser emission from the \transa\/ (6.7-GHz) transition of methanol was
first detected by Menten \shortcite{Me1991b}, who found it to be
common towards star formation regions.  Subsequent observations have
confirmed this, and there are presently more than 250 published sites
of 6.7-GHz methanol emission within the Galaxy
\cite{Ma1992b,Ma1992a,Sc1993,Ca1995a,Ho1995,Va1995b,El1995a}.
Currently all 6.7-GHz methanol masers are thought to be associated
with regions of massive star formation, although the untargeted survey
of Ellingsen \etal\/ \shortcite{El1995a} detected a number of sources
with no known associations.

The 6.7-GHz transition of methanol produces the second strongest
Galactic masers of any molecule.  This makes it ideal for
interferometric observations, as the accuracy with which we can
determine the spatial distribution of maser spots is largely
dependent on the signal-to-noise ratio.  The first high-resolution
spatial images of the 12.2-GHz methanol masers \cite{No1988} showed
that, unlike OH or \water, the methanol masers often exhibit
a simple spatial morphology.  Subsequent observations of the 6.7-GHz
methanol masers in many of the same sources \cite{No1993} revealed that 
they also frequently have a curved or linear spatial structure.

Interferometric observations of the 12.2- and 6.7-GHz methanol masers
show that the methanol masers often emanate from the same region as
the OH masers \cite{Me1988,Me1992,Ca1995e}.  The similarity in the
spectra of many 6.7- and 12.2-GHz methanol masers was first noted by
Menten \shortcite{Me1991b}.  The observations of Menten \etal\/
\shortcite{Me1992} and Norris \etal\/ \shortcite{No1993} show that
when there is a correspondence between one or more 12.2- and 6.7-GHz
spectral features, the maser emission at the two frequencies appears
to originate from the same location.

It has been suggested that 6.7- and 12.2-GHz methanol masers occur in
the circumstellar disc which surrounds massive stars during their
formation \cite{No1993,No1995}.  If we are observing these discs
nearly edge-on, then this provides an explanation for the curved and
linear structures observed in many methanol masers.  One of the
predictions of this model is that the masers should show a simple
velocity structure, and Norris \etal\/ \shortcite{No1995} have
presented evidence for this.  Another prediction is that the masers
should be approximately coincident with the peak of the continuum
emission.  This is in contrast to OH masers, which are typically found
near the edge of \ionhy\/ regions \cite{Ga1987}.  The primary aim of
this paper is to examine whether this prediction is supported by
observation.

\section{Observations}

We have used the ATCA to image the continuum emission of the \ionhy\/
regions associated with the 6.7-GHz methanol masers G318.95-0.20,
G339.88-1.26 and NGC~6334F (G351.42+0.64).  Our observations of
NGC~6334F include three separate clusters of masers, and so subsequent
discussions refer to a total of five clusters of 6.7-GHz methanol
masers.  The observations were made during 1993 November 7 with the
array in the 6A configuration. This has minimum and maximum baseline
lengths of 0.33 and 5.9~km respectively.  The correlator was
configured to record both a 128-MHz band centred at 8.590~GHz, and an
8-MHz band which was alternated between 8.584 and 6.669~GHz.  This
enabled us to make observations of the 6.7-GHz methanol masers and the
\recomb\/ recombination line whilst simultaneously imaging the
continuum emission.  The 8-MHz band was split into 512 channels
yielding a velocity resolution of 0.84 \kms\/ at 6.669~GHz.  The HPBW
(half-power beam width) of the synthesised beam at 8.590~GHz was
approximately 1.2~arcsec.

Each program source was observed twelve times for a 16-min period over
the 13-h observing session and was preceded by a 4-min observation of
a calibration source.  The observing frequency of the 8-MHz band was
changed to the alternate frequency at the end of each observing cycle
(an observation of each program and calibration source).  As a result,
the maser and recombination line observations consist of six 16-min
scans, while the continuum observations consist of 12 scans.


The data were calibrated and imaged using the AT AIPS (Astronomical
Image Processing System), which is based on the NRAO software package
of the same name.  1934-638 was used as the primary flux calibrator,
which we assumed to have flux densities of 3.92~Jy and 2.86~Jy at
6.669~GHz and 8.590~GHz respectively.  1414-59, 1740-517 and 1744-312
were used as secondary calibrators and their positions and flux
densities, calculated by comparing them with 1934-638 are listed in
Table~\ref{tab:sc}.

\begin{table*}
  \caption{Positions and measured flux at 6.7 and 8.5~GHz, of the
           secondary calibrators used}
  \begin{tabular}{lccrr} 
  {\bf Source} & {\bf Right Ascension} & {\bf Declination} & 
    {\bf 6.7-GHz Flux} & {\bf 8.5-GHz Flux} \\
  {\bf Name}   & {\bf (J2000)}         & {\bf (J2000)}     &
    {\bf Density (Jy)} & {\bf Density (Jy)} \\ [2mm]
  \hline\hline 
  $1414\!-\!59$  & 14:17:41.640 & -59:50:37.53 & 1.16 & 1.08 \\
  $1740\!-\!517$ & 17:44:25.454 & -51:44:43.77 & 3.03 & 2.51 \\
  $1744\!-\!312$ & 17:43:59.640 & -31:07:38.45 & 0.40 & 0.40 \\
  \end{tabular} 
  \label{tab:sc}
\end{table*}

After the initial calibration, the continuum emission for each source
was imaged and CLEANed using the AIPS task MX.  The spectral
resolution of our 6.7-GHz methanol maser observations is four times
lower than that of Norris \etal\/ \shortcite{No1993}.  Because of this
we did not try to determine the relative positions of the 6.7-GHz
methanol masers from these data, but rather used the positions
determined by Norris \etal\/ .  We independently
imaged only the reference maser feature.

We used three different techniques to determine the offset between the
position of the 8.5-GHz continuum peak and the reference feature
from the 6.7-GHz methanol maser spectrum.  The first and simplest
technique was to measure independently the absolute positions of the
continuum peak and the reference maser.  The second technique was to
reference the phase of all the channels in the 6.7-GHz methanol maser
observations to a channel containing a strong unresolved maser feature
(the reference feature).  We then formed a continuum image from the
channels which did not contain maser emission, and determined the
position of the peak.  This method works only if the radio continuum
emission from the UC\ionhy\/ region is strong.  The final technique
was to reference the phase of the 8.5-GHz continuum channels to the
6.7-GHz reference feature.  If we assume that the same region of the
atmosphere causes the phase errors at both 6.7 and 8.5~GHz, then we
can correct the 8.5-GHz phase simply by multiplying the 6.7-GHz
correction by the ratio of the frequencies.  Scaling the phase
corrections causes a discontinuity if the reference phase wraps.  For
these observations we found that after calibration the reference phase
did not wrap for many of the baselines.  For those where the phase did
wrap, it did so only once and we flagged these data before phase
referencing the 8.5-GHz data.

To assess the accuracy with which we could measure the offset between
a reference maser feature and the 8.5-GHz continuum peak, we
calculated the offset for NGC~6334F using each of the above methods.
In addition, we calculated the offset for several datasets which had
not been fully calibrated, or contained small deliberate errors in the
calibration.  In all cases the measured offset was very similar, with
the rms being 0.2~arcsec.  The offsets quoted below are the mean of
the offsets calculated from each of the methods outlined above (for
G339.88-1.26 only the first and third methods were used).  We adopt
0.2~arcsec as a conservative estimate of the standard error.

\section{Results}

\begin{figure*}
  \centering
  \begin{minipage}[t]{0.80\textwidth}
    \epsfxsize=0.99\textwidth
    \epsout 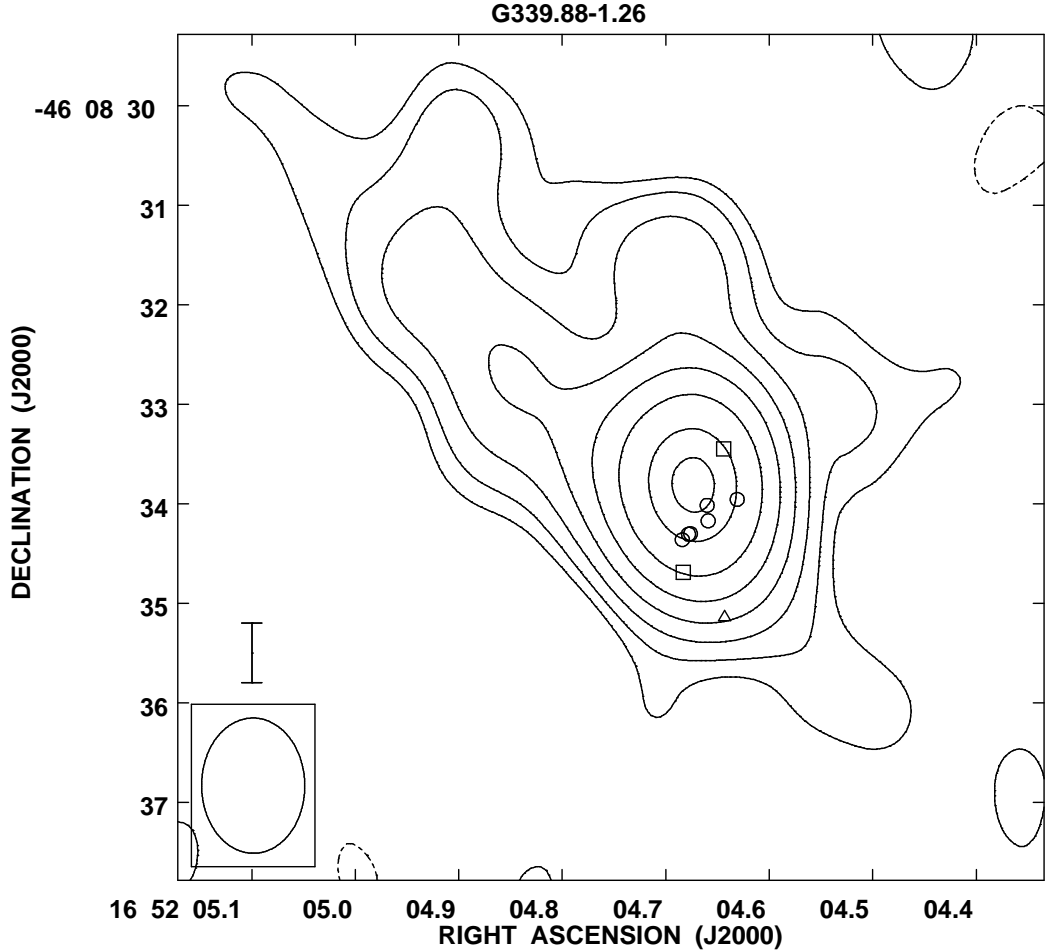
  \end{minipage}
  \caption{Radio continuum image of G339.88-1.26 at 8.59~GHz.  The
           contours are at -1, 1, 2, 4, 8, 16, 32, 64, and 90 per cent
           of the peak flux (6.14 \mjb\/).  The rms noise level in
           the image is approximately 0.04 \mjb\/.  The positions of
           the 6.7-GHz methanol masers as measured by Norris \etal\/
           \protect\shortcite{No1993} are marked as open circles.  The
           position of the OH masers as measured by Caswell, Vaile \&
           Forster \protect\shortcite{Ca1995e} and the \water\/ masers
           as measured by Forster \& Caswell
           \protect\shortcite{Fo1989} are marked as squares and a
           triangle respectively.  The error bar above the clean beam
           plot represents the 3-$\sigma$ error for the relative
           offset between the methanol masers and the continuum.  The
           error in the relative position of the methanol masers with
           respect to each other is smaller than the size of the
           symbols used to plot them.  The width of the synthesised
           beam at the half power points is 1.04 arcsec in Right
           Ascension and 1.36 arcsec in Declination.}
  \label{fig:g339}
\end{figure*}

\begin{figure*}
  \centering
  \begin{minipage}[t]{0.80\textwidth}
    \epsfxsize=0.99\textwidth
    \epsout 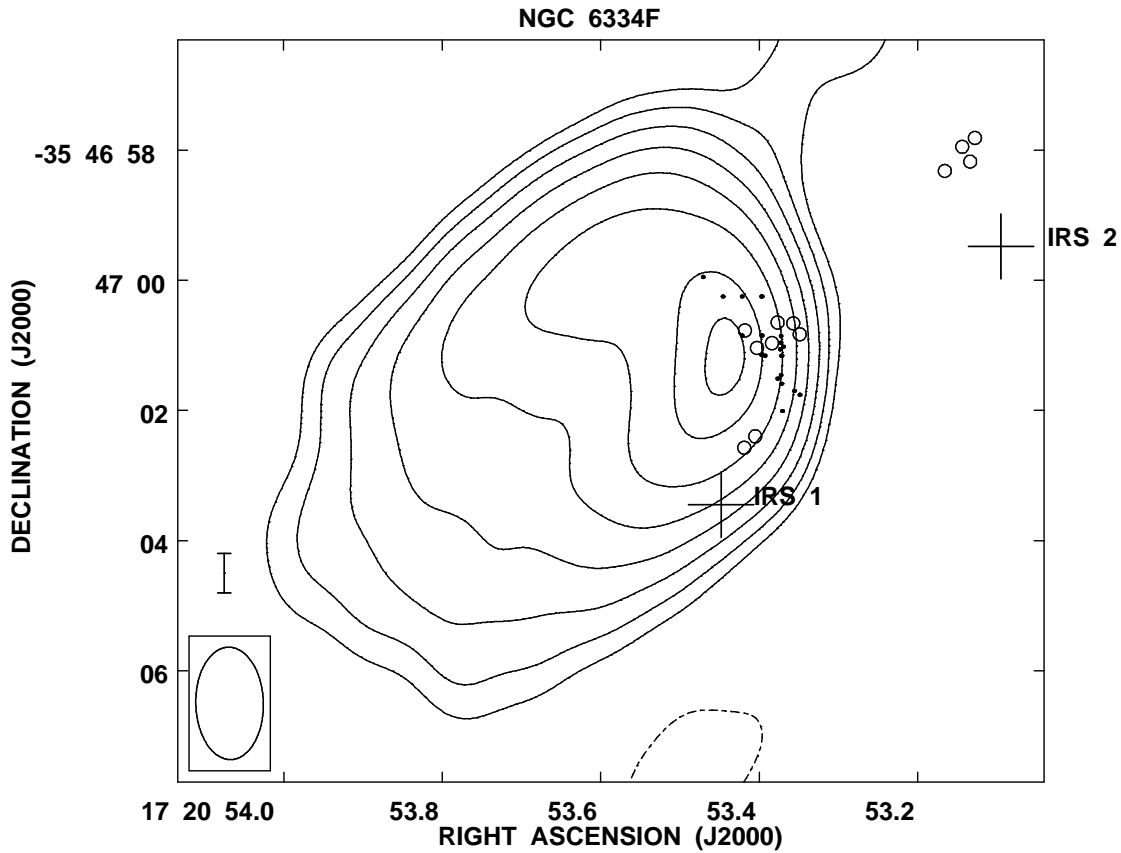
  \end{minipage}
  \caption{Radio continuum image of NGC~6334F at 8.59~GHz.  The
           contours are at -1, 1, 2, 4, 8, 16, 32, 64, and 90 per cent
           of the peak flux (585 \mjb\/).  The rms noise level in the
           image is approximately 1.0 \mjb\/.  The open circles are
           the positions of the 6.7-GHz methanol masers (Norris
           \etal\/ 1993), the dots are the positions of the OH masers
           (Gaume \& Mutel 1987) and the crosses mark the positions of
           two 20-\micron\/ infrared sources (Harvey \& Gatley 1983).
           The error bar above the clean beam plot represents the
           3-$\sigma$ error for the relative offset between the
           methanol masers and the continuum.  The error in the
           relative position of the methanol masers with respect to
           each other is smaller than the size of the symbols used to
           plot them.  The width of the synthesised beam at the
           half-power points is 1.04 arcsec in Right Ascension and
           1.73 arcsec in Declination.}
  \label{fig:ngc6334f}
\end{figure*}

Images of the 8.5-GHz radio continuum with the positions of the
6.7-GHz methanol masers marked are shown in Figs~\ref{fig:g339} and
\ref{fig:ngc6334f}.  Five separate sites of 6.7-GHz methanol maser
emission were observed in the three sources, but we detected radio
continuum emission associated with only two.  The observed parameters
of the radio continuum emission toward each of the sites of maser
emission are summarized in Table~\ref{tab:cont}.  Two of the sites
without continuum emission lie in the NGC~6334 star formation region.
The upper limits on the peak flux density we obtained for them (see
Table~\ref{tab:cont}) are quite large, due to the presence of nearby
strong diffuse sources, which severely limited the dynamic range we
were able to achieve for images of this region.  The resulting upper
limits are not significantly smaller than the peak flux measured for
G339.88-1.26.  The entire NGC~6334 region was imaged with the VLA by
Rodr\'iguez, Cant\'o \& Moran \shortcite{Ro1982}, but they were able
to set only an upper limit of 20~mJy for other compact sources in the
region.  Gaume \& Mutel \shortcite{Ga1987} also imaged NGC~6334F with
the VLA and report no additional compact emission with a 5-$\sigma$
limit of 6.5~mJy for their 15-GHz image.

\begin{table*}
  \caption{Positions of observed sites of 6.7-GHz methanol maser
           emission and some parameters of any associated continuum
           emission.  For the first three sources the positions are
           those of the reference maser in each source.  For
           NGC~6334F (NW) the position is the centroid of the four maser
           spots in that region and for G351.54+0.66 the position is
           that of the -2.5~\kms\/ feature.  The upper limits for
           sources with no detected continuum emission are 5 times the
           rms noise level in the final image.}
  \begin{tabular}{lccccrr}
  {\bf Source} & {\bf Right Ascension} & {\bf Declination} & 
    {\bf 8.5-GHz Peak} & {\bf 8.5-GHz Integrated} & {\bf Major Axis} &
    {\bf Minor Axis} \\
  {\bf Name}   & {\bf (J2000)}         & {\bf (J2000)}     &
    {\bf Flux Density} & {\bf Flux Density}       & {\bf (arcsec)}   &
    {\bf (arcsec)}   \\
               &                       &                   &
    {\bf (mJy)}        & {\bf (mJy)}              &                  &
                     \\ [2mm]
  \hline\hline
  G$318.95\!-\!0.20$ & 15:00:55.332 & -58:58:42.04 & $< 0.82$ & 
    &     &    \\
  G$339.88\!-\!1.26$ & 16:52:01.682 & -46:08:34.42 & 6.14     & 14.0
    & 6.3 & 3.4 \\
  NGC~6334F (C)      & 17:20:53.454 & -35:47:00.52 & 585      & 2780
    & 8.8 & 6.5 \\
  NGC~6334F (NW)     & 17:20:53.240 & -35:46:57.92 & $< 4.8$  &
    &     &     \\
  G$351.54\!+\!0.66$ & 17:20:54.621 & -35:45:07.38 & $< 3.3$  &
    &     &     \\
  \end{tabular}
  \label{tab:cont}
\end{table*}

\subsection{G318.95-0.20}

We are able to set a 5-$\sigma$ upper limit of 0.82~\mjb\/ for any
radio continuum emission from G318.95-0.20.  UC\ionhy\/ regions with
peak flux densities less than our upper limit have been detected.
However, the large-scale, high-resolution studies of UC\ionhy\/
regions which have been made so far
\cite{Wo1989,Ga1993,Ku1994,Mi1994}, typically have sensitivity
comparable to, or worse than, our observation of G318.95-0.20.  Thus
there is little information on whether UC\ionhy\/ regions with peak
fluxes less than 1~mJy are common.

The depth to which we could CLEAN the image toward G318.95-0.20 was
limited by the presence of a confusing source with a peak flux density
of 29.5~\mjb\/ $\sim$ 160~arcsec west and 34.5~arcsec north of the
reference maser position.  The confusing source we detected is
approximately coincident with the {\em IRAS} source 14567-5846, which
has been identified as an \ionhy\/ region.  The radio continuum
emission appears to be a cometary \ionhy\/ region with a major axis of
$\sim$ 10~arcsec.  Two groups have reported maser emission toward
14567-5846 \cite{Co1988,Ke1988}.  In both cases they report the
position to be consistent with the OH maser position measured by
Caswell \& Haynes \shortcite{Ca1987}, which, in turn, is coincident
with the 6.7-GHz \methanol\/ masers.  Therefore, it appears that all
the maser emission reported in this region to date is from the same
site, which is more than 2 arcmin away from the {\em IRAS} source
14567-5846.

\subsection{G339.88-1.26}

Our 8.5-GHz image of G339.88-1.26 is shown in Fig.~\ref{fig:g339}.
This is the first image to be produced of this UC\ionhy\/ region.  We
measured it to have a peak brightness of 6.1~\mjb\/ at 8.5~GHz, but
it was too weak to image from our 6.7-GHz spectral-line observations.
The bulk of the continuum emission is unresolved in our 1.2-arcsec
synthesised beam, but shows some low-level extension to the north-east,
suggesting a possible cometary morphology.

The reference maser (-38.7~\kms\/) is the most south-eastern of the
spots and is offset from the continuum peak by 0.6$\pm$0.2~arcsec.
The methanol masers lie in a line approximately across the centre of
the continuum emission, perpendicular to the direction of the extended
emission.  The positions of two of the OH maser spots have been
observed by Caswell \etal\/ \shortcite{Ca1995e}, and they straddle the
line of 6.7-GHz methanol masers.  The position of one of the \water\/
masers spots was measured with the VLA by Forster \& Caswell
\shortcite{Fo1989}.  They quote an absolute positional accuracy of
0.5~arcsec, but as G339.88-.126 was the most southerly source in their
sample, the maser position is probably less well determined than for
the majority of sources.  The \water\/ maser positions they quote is
approximately 1~arcsec south of the 6.7-GHz methanol masers.

\subsection{NGC~6334F (G351.42+0.64)}


This source was previously imaged at 4.9 and 15~GHz using the VLA
\cite{Ro1982,Ga1987}.  Our 8.5-GHz image, shown in
Fig.~\ref{fig:ngc6334f}, agrees with theirs.  Making sensitive
high-resolution images of the NGC~6334F region is difficult because of
the presence of the sources NGC~6334D and E, two nearby strong,
diffuse \ionhy\/ regions.  We measure a peak brightness of 585~\mjb\/
for NGC~6334F at 8.5~GHz and 606~\mjb\/ at 6.7~GHz.  We also produced
an 8.5-GHz image using a restoring beam with the same dimensions as
the 6.7-GHz beam and measured the spectral index ($\alpha$) of the
peak to be 0.95 ($S_{\nu} \propto \nu^{\alpha}$) between 6.7 and
8.5~GHz.  This implies that the centre of the \ionhy\/ region is still
optically thick at 8.5~GHz.

Toward NGC~6334F, three centres of 6.7-GHz methanol maser emission are
within the primary beam of the ATCA antennas.  We have labelled the
three sites NGC~6334F (C) (all those masers which lie in projection
against the \ionhy\/ region), NGC~6334F (NW) (the masers to the
north-west of NGC~6334F (C)) and G351.54+0.66.  G351.54+0.66 is not
shown in Fig.~\ref{fig:ngc6334f}, as it is offset 14.0~arcsec east and
114.2~arcsec north of NGC~6334F (C) (see Table~\ref{tab:cont}).  The
northern clump of masers in NGC~6334F (C) is approximately coincident
with the position of the OH masers determined by Gaume \& Mutel
\shortcite{Ga1987}.  No OH masers have been detected at the locations
of either NGC~6334F (NW) or G351.54+0.66.  The latter is in the same
general region as the infrared source NGC~6334I(N), but does not
appear to be closely associated with any known radio or infrared
sources.

\section{Discussion}


\subsection{The Energizing star}

By making some simplifying assumptions about the nature of the
\ionhy\/ regions observed, we can calculate some of the physical
parameters (see Table~\ref{tab:hii}).  We calculated the electron
density (n$_{e}$), emission measure (EM) and mass of ionized hydrogen
(M$_{H\sc{II}}$) using the equations of Panagia \& Walmsley
\shortcite{Pa1978}.  These equations assume that the \ionhy\/ region
is optically thin, spherically symmetric and has an electron
temperature of 10$^{4}$~K.  In neither case are the \ionhy\/ regions
we observe spherically symmetric.  Instead, we calculate the linear
radius of the \ionhy\/ region using the method outlined in Panagia \&
Walmsley independently for Right Ascension and Declination and use the
geometric mean of the two.  We also calculate the excitation parameter
using the formula of Schraml \& Mezger \shortcite{Sc1969}.  Using this
we estimate the Lyman continuum photon flux (N$_{L}$) and the spectral
class of the exciting star \cite{Pa1973}.  As the excitation parameter
(U) depends only upon the flux density and the distance to the source,
we were also able to calculate upper limits for the spectral class of
the exciting star where we did not detect any continuum emission.  For
NGC~6334F, the spectral index we calculate for the central peak of the
source between 6.7 and 8.5~GHz indicates that it is optically thick at
these frequencies.  The equations we have used to calculate the
\ionhy\/ region parameters assume that they are optically thin.  The
effect of the violation of this assumption is to reduce our estimate
of the ionizing flux of the exciting star, effectively making it a
lower limit.  So for the \ionhy\/ region NGC6334F, which has a central
core that appears to be optically thick from our spectral index
calculations, the stellar types we derive are a lower limit.  However,
the effect is small since Gaume \& Mutel \shortcite{Ga1987} derive
similar parameters from observations at 15~GHz where the optical depth
effect will be much smaller.

\begin{table*}
  \caption{Derived parameters for \ionhy\/ regions and the exciting stars
           associated with them.  The distance estimates are from :
           1~=~Caswell \& Haynes \protect\shortcite{Ca1987},
           2~=~Caswell \& Haynes \protect\shortcite{Ca1983},
	   3~=~Neckel \protect\shortcite{Ne1978}}
  \begin{tabular}{lrrrrrrr}
  {\bf Source} & {\bf Distance} & {\bf n$_{e}$}     & {\bf EM}             & 
    {\bf M$_{H{\sevensize II}}$} & {\bf U}             & {\bf Log N$_{L}$} &
    {\bf Spectral} \\
               & {\bf (kpc)}    & {\bf (cm$^{-3}$)} & {\bf (pc cm$^{-6}$)} &
    {\bf (M\sun)}        & {\bf (pc cm$^{-2}$)} & {\bf (s$^{-1}$)}   &
    {\bf Type} \\
  \hline\hline
  G$318.95\!-\!0.20$ & 2.0$^{1}$
    &              &              &       & $<$2.1 & $<$44.73 & $<$B2   \\  
  G$339.88\!-\!1.26$ & 3.0$^{2}$
    & 2.6x10$^{4}$ & 5.9x10$^{7}$ & 0.001 &    7.2 &    46.32 &    B0.5 \\
  NGC~6334F (C)      & 1.7$^{3}$
    & 9.5x10$^{4}$ & 9.6x10$^{7}$ & 0.025 &   25.7 &    47.98 &    O9   \\
  NGC~6334F (NW)     & 1.7$^{3}$ 
    &              &              &       & $<$3.5 & $<$45.36 & $<$B1   \\
  G$351.54\!+\!0.66$ & 1.7$^{3}$ 
    &              &              &       & $<$3.1 & $<$45.20 & $<$B1   \\
  \end{tabular}
  \label{tab:hii}
\end{table*}

\subsection{Morphology}

Norris \etal\/ \shortcite{No1993} observed 15 sites of 6.7-GHz
methanol maser emission and found that a large fraction of their
sample of 6.7- and 12.2-GHz methanol maser sources have a simple
curved or linear spatial distribution.  Based on their spatial
distribution, we can divide all 6.7-GHz methanol maser sources into one
of two classes: those with a simple linear or curved morphology
(e.g. G339.88-1.26), and those with a more complex morphology
(e.g. NGC~6334F).  We have radio continuum observations associated
with only two sites of 6.7-GHz methanol maser emission, one from each
class of spatial morphology.  No models have been suggested to explain
the methanol masers with complex spatial distributions.  They may
represent a different evolutionary phase of the star formation
process, as suggested by Forster \& Caswell \shortcite{Fo1989} for
complex OH and \water\/ maser distributions.  Three possibilities have
been suggested to explain the curved/linear spatial morphology :
shocks fronts, collimated jets, and circumstellar discs
\cite{No1993,No1995}.  For each of these, Norris \etal\/
\shortcite{No1995} make a specific prediction about the relationship
between the methanol masers and the \ionhy\/ region.  If the methanol
masers form in the shocks at the interface between the ionized and
molecular gas, then we expect to see them near the edge of \ionhy\/
regions, as is often observed for OH.  If the methanol masers form in
jets or collimated outflows, then we would expect to observe them
distributed radially to the \ionhy\/ region.  However, if the methanol
masers form in the circumstellar discs of young stars, then
we would expect to observe them approximately coincident with the
continuum peak for the \ionhy\/ region.

\subsubsection{G339.88-1.26} \label{sec:g339}

G339.88-1.26 is one of the strongest sources of both 6.7- and 12.2-GHz
methanol maser emission.  The only published high-resolution infrared
observations of the G339.88-1.26 region \cite{Te1994} show a strong
peak, which appears to be slightly offset from the position we measure
for the UC\ionhy\/ region.  The methanol maser emission has an
approximately linear spatial distribution at both 6.7 and 12.2~GHz
\cite{No1988,No1993}, and we measure the masers to be offset slightly
to the south-west of the peak in the continuum emission, but not
significantly so.  This is the position predicted for the 6.7-GHz
methanol masers if they occur in circumstellar discs.

Unlike the general case observed for OH masers, methanol masers do not
lie toward the edge of the \ionhy\/ region \cite{Ga1987}, which seems
to rule out the hypothesis that the masers emanate from shocked gas.
Conceivably these observations are compatible with the masers lying in
a highly collimated outflow, as they are radial to the \ionhy\/
region.  However, the distribution of the masers is highly linear and
they do not have a wide velocity range as might be expected if they
emanated from a high-velocity outflow.  Further, we also require a
double-sided jet, which is almost never seen in stars.

From our radio continuum images we estimate that a B0.5 star is
required to produce the observed \ionhy\/ region.  However, as the
model we used does not take into account the absorption of UV photons
by dust, this is a lower limit on the spectral class of the exciting
star.  Norris \etal\/ \shortcite{No1995} used simple modelling to show
that the observed spatial and velocity distribution of the masers is
consistent with Keplerian motion, but were unable to derive a mass for
the star.  We find that the position of the masers with respect to the
parent \ionhy\/ region also supports the hypothesis that the masers
lie in circumstellar discs.

\subsubsection{NGC~6334F} \label{sec:ngc6334f}

The NGC~6334 region is the most active known region of OB star
formation in the Galaxy \cite{Ha1983}.  The region has been the
subject of several surveys at radio and infrared wavelengths
\cite{Mc1979,Ro1982,Lo1986,St1989a}.  These observations show six main
sites of radio emission and a similar number of clusters of infrared
emission.  Unfortunately, the nomenclature of the region is rather
confusing.  The main site of 6.7-GHz methanol maser emission, called
G351.42+0.64 in the methanol maser literature, is associated with the
radio source NGC~6334F, which is designated NGC~6334-I at infrared
wavelengths.

The NGC~6334F (C) star formation region is often compared to W3(OH).
Both these regions show maser emission in many molecular transitions
which are rarely detected toward other \ionhy\/ regions.  For example,
very strong maser emission from many class II methanol transitions,
along with thermal emission from class I transitions, has been
detected toward NGC~6334F (C)
\cite{Ba1987,Ha1989a,Ha1989b,Me1989,Me1991b}.  This suggests that
W3(OH) and NGC~6334F (C) have some special characteristics not
generally shared with other massive star-formation regions.  Therefore
it may not be valid to infer general characteristics of molecular
emission, or star formation, from observations of NGC~6334F (C) or
W3(OH) alone.

It is well established that 6.7-GHz methanol maser emission is common
toward star formation regions which show main-line OH maser emission
\cite{Me1991b,Ma1992b,Ga1994,Ca1995a}.  However, recent observations
by Ellingsen \etal\/ \shortcite{El1995a} have detected many new
6.7-GHz methanol masers in a region of the Galactic Plane previously
surveyed for OH maser emission by Caswell, Haynes \& Goss
\shortcite{Ca1980}.  This suggests that in some sources the conditions
are favourable for methanol maser emission, but not OH.  Menten
\etal\/ \shortcite{Me1992} found OH masers situated in, or near, all
but one of the clumps of methanol maser emission in W3(OH).  However,
within the clumps there appears to be an anticorrelation between the
positions of the OH and methanol masers.  For NGC~6334F (C) we find,
in agreement with Menten \etal\/, that the OH masers appear to be
associated with methanol maser emission.  However, unlike the case for
W3(OH) there are two clumps of 6.7-GHz methanol maser emission which
have no reported OH emission.  This, combined with the large scale
observations of Ellingsen \etal\/ \shortcite{El1995a}, suggests that
6.7-GHz methanol masers are more widespread than OH masers in star
formation regions.

Infrared observations at a large number of wavelengths by Harvey \&
Gatley \shortcite{Ha1983} found the strongest emission in all bands to
be coincident with the radio continuum peak (which they designated
IRS~1).  At 20 and 30~\micron\/ the emission is extended to the
North-west indicating the presence of a second source (IRS~2)
separated from IRS~1 by approximately 6~arcsec.  They suggest that
IRS~1 and 2 may form a double source.  2.2-\micron\/ (K-band) images
of the same region by Straw \etal\/ \shortcite{St1989a} detected a
cluster of sources in the general region of NGC~6334F, including a
counterpart to IRS~1, but no counterpart to IRS~2.

Observations of CO, \ammonia\/ and recombination lines probe
large-scale outflows and the velocity of the ionized gas, and so
provide additional information on star formation regions.  Two lobes
of \ammonia\/ and CO emission have been observed to the north-east and
south-west of the radio continuum \cite{Ja1988,Ba1990}.  The lobes are
approximately perpendicular to the major-axis of the continuum
emission and both show a velocity gradient.  A rotating molecular disc
around IRS~1 \cite{Ja1988} and bipolar outflow from IRS~1
\cite{Ba1990} have been proposed as explanations for the observed
distribution of the molecular gas.  The detection of shock-excited
\molhy\/ in the general vicinity of NGC~6334F \cite{St1989b} and
shock-excited \ammonia\/ (3,3) masers at the interface between the
molecular gas and ambient medium \cite{Kr1995}, support the outflow
hypothesis.  Recent recombination-line observations also show a
velocity gradient in the ionized gas, in approximately the same
directions as the molecular gas, but are inconsistent with molecular
outflow from IRS~1 \cite{De1995}.  De Pree \etal\/ suggest instead
that IRS~2 may be the source of the observed molecular outflow

The orientation of IRS~1 and 2 is in the same direction, and the
separation roughly the same distance, as the two main clusters of
6.7-GHz methanol masers near NGC~6334F which we have labelled
NGC~6334F (C) and (NW).  Norris \etal\/ \shortcite{No1995} have shown
that the observed spatial and velocity distribution of the two
clusters of methanol masers is consistent with Keplerian motion about
two separate stars.  If IRS~2 is the source of the molecular outflow
then we would expect it to have a circumstellar disc perpendicular to
the direction of the outflow.  The 6.7-GHz methanol masers NGC~6334F (NW)
have a spatial and velocity structure consistent with Keplerian motion
and are aligned perpendicular to the molecular outflow. This leads us
to speculate that NGC~6334F (NW) may be associated with the infrared
source IRS~2.  If we can show that IRS~2 is coincident with the
masers, and that it is the source of the molecular outflow, then this
source provides circumstantial evidence for the circumstellar disc
model of methanol masers.  This requires higher resolution
observations in the mid- to far-infrared and of the molecular outflow,
particularly at its origin.

In Table~\ref{tab:hii} we present some calculated physical parameters
of the \ionhy\/ regions and exciting stars.  Our values are all
comparable with those calculated by Rodr\'iguez \etal\/
\shortcite{Ro1982} and Gaume \& Mutel \shortcite{Ga1987}.  From their
model of the 6.7-GHz methanol masers, Norris \etal\/
\shortcite{No1995} calculate a mass of 70~M\sun\/ for IRS~1 and
13~M\sun\/ for IRS~2.  These masses are approximate but, encouragingly
both are roughly consistent with the spectral types inferred from the
radio continuum observations.

\section{Conclusion}

We have detected 8.5-GHz continuum emission toward two of the five
sites of 6.7-GHz methanol maser emission detected.  For G339.88-1.26
we find the position of the 6.7-GHz methanol masers to be consistent
with the hypothesis that these masers occur in circumstellar discs of
massive stars.  The maser emission toward NGC~6334F has a more complex
distribution, with one of the clusters of methanol maser emission
being approximately coincident with the OH emission detected by Gaume
\& Mutel \shortcite{Ga1987}.  We also argue that the 6.7-GHz methanol
maser cluster we call NGC~6334F (NW) may be in a circumstellar disc of
the young stellar object IRS~2 \cite{Ha1983}.

\section*{Acknowledgments}

This research has made use of the Simbad database, operated at CDS, 
Strasbourg, France.

\label{lastpage}

\end{document}